\documentclass [11,onecolumn]{mn2e}

\title[Improved equations for HTS]{Improved equations for eccentricity generation in
hierarchical triple systems}
\author[Nikolaos Georgakarakos]{Nikolaos Georgakarakos\\ATEI of Kalamata,
\\Department of Technology of Informatics and Telecommunications, \\
7 Kiekies Str., Sparta 23100,Greece\\
email: georgakarakos@hotmail.com}
\date{}

\usepackage{graphicx}

\begin{document}
\maketitle
\begin{abstract}
In a series of papers, we developed a technique for estimating the
inner eccentricity in hierarchical triple systems, with the inner
orbit being initially circular.  However, for certain combinations
of the masses and the orbital elements, the secular part of the
solution failed. In the present paper, we derive a new solution
for the secular part of the inner eccentricity, which corrects the
previous weakness. The derivation applies to hierarchical triple
systems with coplanar and initially circular orbits.  The new
formula is tested numerically by integrating the full equations of
motion for systems with mass ratios from ${10^{-3} \hspace{0.2cm}
\mbox{to} \hspace{0.2cm} 10^{3}}$. We also present more numerical
results for short term eccentricity evolution, in order to get a
better picture of the behaviour of the inner eccentricity.
\end{abstract}

\noindent {\bf Key words:} Celestial mechanics, binaries:general,
binaries:close.

\section{INTRODUCTION}
\label{s1} Generally, stars have a tendency to form groups of
different multiplicity, from the smallest possible (binary
systems) up to large groups, like globular clusters with a
population of the order of ${10^{7}}$ stars. Observations give
values for the frequency of multiple stars in the galactic field
of up to
\begin{math}
70 \%,
\end{math}
(Kiseleva-Eggleton ${\&}$ Eggleton 2001 and references therein). Most
of the multiple systems seem to be hierarchical (e.g. Tokovinin
1997, Kiseleva-Eggleton ${\&}$ Eggleton 2001).

A hierarchical triple system consists of a binary system and a
third body on a wider orbit.  The motion of such a system can be
pictured as the motion of two binaries: the binary itself (inner
binary) and the binary which consists of the third body and the
centre of mass of the inner binary (outer binary). For most
hierarchical triple stars, the period ratio ${X}$ is of the order
of 100 and these systems are probably very stable dynamically.
However, there are systems with much smaller period ratios, like
the system HD 109648 with ${X=22}$ (Jha et al. 2000), the
${\lambda}$ Tau system, with
\begin{math}
X=8.3
\end{math}
(Fekel ${\&}$ Tomkin 1982) and the CH Cyg system with
\begin{math}
X=7.0
\end{math}
(Hinkle et al. 1993).

When the inner pair of a hierarchical  triple stellar system is a
 close binary system, i.e. the separation between the components
is comparable to the radii of the bodies, the evolution of the
inner binary can depend very sensitively on the separation of its
components and this in turn is affected by the third body. Thus, a
slight change in the separation of the binary stars can cause
drastic changes in processes such as tidal friction and
dissipation, mass transfer, accretion and  mass loss due to a
stellar wind, which may result in changes in stellar structure and
evolution.   For instance, an eccentricity of only ${\sim 0.001}$
can be important in a semidetached binary (Kisela, Eggleton ${\&}$
Mikkola 1998). A comprehensive summary of those topics can be found
in Eggleton (2006).

In a close binary, the orbit is expected to become circular
eventually, due to tidal friction. However this is not possible
when the binary motion is affected by the perturbations of a third
companion (Mazeh ${\&}$ Shaham 1979, Mazeh 2008).
Eggleton and Kiseleva (1996), based on results from numerical
integrations of coplanar, prograde and initially circular orbits
of hierarchical triple systems, derived the following empirical
formula for the mean inner eccentricity:
\begin{equation}
\bar{e}_{{\rm in}}=\frac{A}{X^{1.5}\sqrt{X-B}}, \label{fegg}
\end{equation}
where ${A}$ and ${B}$ depend on the mass ratios.

Equation (\ref{fegg}) was confirmed analytically by Georgakarakos
(2002). Moreover, the analytic derivation  was generalized to
hierarchical triple stellar systems with other orbital
characteristics (Georgakarakos 2003 and 2004).  Subsequently, some
of the formulae were also tested for systems with planetary mass
ratios (Georgakarakos 2006).

In the present paper, we deal with a problem that had to do with
the secular part of our analytical solution and we manage to
obtain a new formula which is free of that problem and gives
improved results for certain cases. The new derivation applies to
systems with initially circular and coplanar orbits, i.e. the type
of systems we dealt with in Georgakarakos (2002).  The new formula
is tested numerically for systems with stellar and planetary mass
ratios.

Other recent work on the dynamics of hierarchical triple systems
includes the work done by Krymolowski ${\&}$ Mazeh (1999), Ford,
Kozinsky ${\&}$ Rasio (2000), Blaes, Lee ${\&}$ Socrates (2002) and
Lee ${\&}$ Peale (2003).

\section{THEORY}
\label{sss}

First, we would like to give a brief summary describing the
derivation of the analytic formula (for more details, one can
check Georgakarakos 2002). Our eccentricity expression consisted
of two parts: a short period part, which varied on a time-scale
comparable to the inner and outer orbital period, and a secular
part.

In order to obtain the short period part of the inner
eccentricity, and by using Jacobian coordinates, we expanded the
perturbing potential in terms of Legendre polynomials and we then
used the definition of the Runge-Lenz vector  to obtain a series
expansion for the inner eccentricity in powers of ${1/X}$ (the
period ratio X was considered to be rather large for the
applications discussed).

The secular part of the eccentricity was derived by using a
Hamiltonian which was averaged over the short period time-scales
by means of the Von Zeipel method.  Then, by manipulating the
canonical equations and by making a few approximations, we
obtained our secular solution and eventually, we combined the
short period and the secular solution in order to get a unified
expression for the inner eccentricity (we also had to find an
expression for the outer eccentric vector, in order to determine
the required initial conditions for the secular problem).

However, the secular solution was singular for certain
combinations of the orbital elements and masses of some systems,
and as a result, our model failed to describe the eccentricity
evolution for those systems.  The problem originated from the
approximations we made about the motion of the outer pericentre,
but with a more careful handling, that can now be corrected (see
subsection \ref{s3}).

Finally, we would like to mention that an important aspect of the
theory that was developed, was the combination of the short period
and secular terms in the expressions for the eccentricities.  At
any moment of the evolution of the system, we considered that the
eccentricity (inner or outer) consisted of a short period and a
long period (secular) component, i.e. ${e=e_{{\rm short}}+e_{{\rm
sec}}}$ (one can picture this by recalling the expansion of the
disturbing function in solar system dynamics, where the perturbing
potential is given as a sum of an infinite number of cosines of
various frequencies). Thus, considering the eccentricity to be
initially zero leads to ${e_{{\rm short}}=-e_{{\rm sec}}}$
(initially), which implies that, although the eccentricity is
initially zero, the short period and secular eccentricity may not
be.

We would like to point out here, that our derivation was done in
the context of the gravitational non-relativistic three body
problem with point masses. However, whenever it is necessary,
it would be possible to
generalize the calculation to include relativistic effects and the
effect of treating the bodies as non-point masses, by adding the
required terms (e.g. a post-Newtonian correction) in the equation
of motion of the inner binary.

\subsection{Short period eccentricity} \label{s2}

First, we list the components of both the inner and outer
eccentric vectors (short period). Although the derivation details
can be found in Georgakarakos (2002), the inner eccentric vector
equations in that paper contain only the two dominant terms of the
expansion, instead of the four we give below (this is done for
better accuracy purposes; a four term expansion was used in
Georgakarakos 2003 and 2004). The expansion of the outer eccentric
vector includes one more term, compared to what we found in
Georgakarakos (2002). That extra term improves the general
solution of the problem, as it provides better initial conditions
for the secular problem.

The components of the inner eccentric vector are:

\begin{eqnarray}
e_{11}(t) & = & x_{1}(t)+C_{\rm
e_{11}}=\frac{m_{3}}{M}\frac{1}{X^{2}}\left( P_{{\rm
x21}}(t)+\frac{1}{X}P_{{\rm x22}}(t)+m_{*}X^{\frac{1}{3}}P_{{\rm
x31}}(t)+m_{*}\frac{1}{X^{\frac{2}{3}}}P_{{\rm
x32}}(t)\right)+C_{\rm e_{11}}
\label{e11}\\
e_{12}(t) & = & y_{1}(t)+C_{\rm e_{12}}=
\frac{m_{3}}{M}\frac{1}{X^{2}}\left(P_{{\rm
y21}}(t)+\frac{1}{X}P_{{\rm y22}}(t)+m_{*}X^{\frac{1}{3}}P_{{\rm
y31}}(t)+m_{*}\frac{1}{X^{\frac{2}{3}}}P_{{\rm
y32}}(t)\right)+C_{\rm e_{12}}  \label{e12}
\end{eqnarray}
where
\begin{eqnarray}
P_{{\rm x21}}(t) & = &
-\frac{1}{2}\cos{n_{1}t}+\frac{1}{4}\cos{((3n_{1}-2n_{2})t-2\phi)}+\frac{9}{4}\cos{((n_{1}-2n_{2})t-2\phi)}\nonumber\\
P_{{\rm x22}}(t) & = & \frac{1}{6}\cos{((3n_{1}-2n_{2})t-2\phi)}+\frac{9}{2}\cos{((n_{1}-2n_{2})t-2\phi)}\nonumber\\
P_{{\rm x31}}(t) & = & \frac{15}{16}\cos{(n_{2}t+\phi)}\nonumber\\
P_{{\rm x32}}(t) & = &
\frac{3}{32}\cos{((2n_{1}-n_{2})t-\phi)}-\frac{45}{32}\cos{((2n_{1}-3n_{2})t-3\phi)}-
\frac{15}{64}\cos{((4n_{1}-3n_{2})t-3\phi)}\nonumber\\
P_{{\rm y21}}(t) & = & -\frac{1}{2}\sin{n_{1}t}+\frac{1}{4}\sin{((3n_{1}-2n_{2})t-2\phi)}-\frac{9}{4}
\sin{((n_{1}-2n_{2})t-2\phi)}\nonumber\\
P_{{\rm y22}}(t) & = & \frac{1}{6}\sin{((3n_{1}-2n_{2})t-2\phi)}-\frac{9}{2}\sin{((n_{1}-2n_{2})t-2\phi)}\nonumber\\
P_{{\rm y31}}(t) & = & \frac{15}{16}\sin{(n_{2}t+\phi)}\nonumber\\
P_{{\rm y32}}(t) & = &
\frac{3}{32}\sin{((2n_{1}-n_{2})t-\phi)}+\frac{45}{32}\sin{((2n_{1}-3n_{2})t-3\phi)}-
\frac{15}{64}\sin{((4n_{1}-3n_{2})t-3\phi)}\nonumber
\end{eqnarray}
and
\begin{displaymath}
m_{*}=\frac{m_{2}-m_{1}}{(m_{1}+m_{2})^{\frac{2}{3}}M^{\frac{1}{3}}}.
\end{displaymath}

The components of the outer eccentric vector, which are obtained
in a similar way to the one we used to get the components of the
inner eccentric vector, are:

\begin{eqnarray}
e_{21}(t) & = & x_{2}(t)+C_{\rm
e_{21}}=\frac{m_{1}m_{2}}{(m_{1}+m_{2})^{\frac{4}{3}}M^{\frac{2}{3}}}\frac{1}{X^{\frac{4}{3}}}[
\frac{3}{4}\cos{(n_{2}t+\phi)}+\frac{1}{X}
(-\frac{3}{16}\cos{((2n_{1}-n_{2})t-\phi)}-\nonumber\\
& & -\frac{21}{16}\cos{((2n_{1}-3n_{2})t-3\phi)})]+C_{\rm e_{21}}\label{e21}\\
e_{22}(t) & = & y_{2}(t)+C_{\rm
e_{22}}=\frac{m_{1}m_{2}}{(m_{1}+m_{2})^{\frac{4}{3}}M^{\frac{2}{3}}}\frac{1}{X^{\frac{4}{3}}}[
\frac{3}{4}\sin{(n_{2}t+\phi)}+\frac{1}{X}
(-\frac{3}{16}\sin{((2n_{1}-n_{2})t-\phi)}+\nonumber\\
& & +\frac{21}{16}\cos{((2n_{1}-3n_{2})t-3\phi)})]+C_{\rm
e_{22}}\label{e22}.
\end{eqnarray}

${M}$ is the total mass of the system, ${n_{1}}$ and ${n_{2}}$ are
the mean motions of the inner and outer binary respectively,
${\phi}$ is the initial relative phase of the two binaries, i.e.
the initial angle between the two Jacobi vectors ${\bmath {r}}$
and ${\bmath {R}}$,  and finally, ${C_{{\rm e}_{11}}}$,${C_{{\rm
e}_{12}}}$, ${C_{{\rm e}_{21}}}$ and ${C_{{\rm e}_{22}}}$ are
constants of integration.

We would like to mention here, that when we expanded the
components of the eccentric vector as a power series in terms of
${1/X}$ , terms of the same order as ${P_{22}}$ were missed out.
Those terms arose from the ${P_{5}}$ Legendre polynomial in the
expansion of the perturbing potential and they are of the form
\begin{displaymath}
P_{51}\sim \frac{105}{128}\frac{m_{3}}{M}\frac{1}{X^{2}}
\frac{1}{X}\frac{1}{M}\frac{m^{4}_{2}-m^{4}_{1}}{(m_{1}+m_{2})^{3}}.
\end{displaymath}
In the best case scenario, a rough estimate gives
\begin{displaymath}
P_{22}/P_{51} \approx 6,
\end{displaymath}
which leads us to conclude that the impact of the ${P_{51}}$ terms
would not be significant for the parameter ranges discussed in
this paper. This will be confirmed by numerical integrations in
the subsequent sections. For that reason, and in order to avoid
further increase of the volume of the equations, we choose not to
include those terms here.

\subsection{Improved secular solution for the eccentricity}
\label{s3}

As in our previous papers, and in order to describe the long term
motion of the system, we use a Hamiltonian which is averaged over
the inner and outer orbital periods by means of the Von Zeipel
method.  There are also other methods of studying the secular
behaviour of hierarchical triple systems, e.g. see Libert ${\&}$
Henrard (2005), Migaszewski ${\&}$ Gozdziewski (2008), both
concentrating on planetary systems. However, those methods cannot
reproduce fully the results of the Von Zeipel averaging method (we
are referring to terms that arise from the canonical
transformation and which are second order in the strength of the perturbation).  
As a result, the outcome is similar for
planetary mass ratios, but for systems with comparable masses, the
extra terms that the Von Zeipel method produces can have a noticeable effect
(e.g. see Krymolowski and Mazeh 1999).

 The doubly averaged Hamiltonian for coplanar orbits is (Marchal
1990):

\begin{eqnarray}
H & = &-\frac{Gm_{1}m_{2}}{2a_{{\rm
S}}}-\frac{G(m_{1}+m_{2})m_{3}}{2a_{{\rm T}}}+Q_{1}+Q_{2}+Q_{3}, \label{hamilto} \\
\mbox{where}\nonumber\\
Q_{1} & = & -\frac{1}{8}\frac{Gm_{1}m_{2}m_{3}a^{2}_{{\rm
S}}}{(m_{1}+m_{2})a^{3}_{{\rm T}}(1-e^{2}_{{\rm T}})^{\frac{3}{2}}}(2+3e^{2}_{{\rm S}}), \\
Q_{2} & = & \frac{15Gm_{1}m_{2}m_{3}(m_{1}-m_{2})a^{3}_{{\rm
S}}e_{{\rm S}}e_{{\rm
T}}}{64(m_{1}+m_{2})^{2}a^{4}_{{\rm T}}(1-e^{2}_{{\rm T}})^{\frac{5}{2}}}\cos{(g_{{\rm S}}-g_{{\rm T}})}(4+3e^{2}_{{\rm S}}),\label{marq}\\
Q_{3} & = & -\frac{15}{64}\frac{Gm_{1}m_{2}m_{3}^{2}a_{{\rm
S}}^{\frac{7}{2}}e_{{\rm S}}^{2}(1-e_{{\rm
S}}^{2})^{\frac{1}{2}}}{(m_{1}+m_{2})^{\frac{3}{2}}M^{\frac{1}{2}}a_{{\rm
T}}^{\frac{9}{2}}(1-e_{{\rm T}}^{2})^{3}}[5(3+2e_{{\rm
T}}^{2})+3e^{2}_{{\rm T}}\cos{2(g_{{\rm S}}-g_{{\rm T}})}].
\end{eqnarray}

The subscripts S and T refer to the inner and outer long period
orbit respectively, while ${g}$ is used to denote longitude of
pericentre.  The first term in the Hamiltonian is the Keplerian
energy of the inner binary, the second term is the Keplerian
energy of the outer binary, while the other three terms represent
the interaction between the two binaries.  The ${Q_{1}}$ term
comes from the ${P_{2}}$ Legendre polynomial, the ${Q_{2}}$ term
comes from the ${P_{3}}$ Legendre polynomial and the ${Q_{3}}$
term arises from the canonical transformation.  The expression for
${Q_{3}}$ in Marchal 1990, contains just the term which is
independent of the pericentres. The complete expression we use
here is taken from Krymolowski and Mazeh (1999).

Using the above Hamiltonian, the secular equations of motion are
the following:

\begin{eqnarray}
\frac{{\rm d}x_{{\rm S}}}{{\rm d}\tau} & = &
\frac{5}{16}\alpha\frac{e_{{\rm T}}}{(1-e^{2}_{{\rm
T}})^\frac{5}{2}}(1-e^{2}_{{\rm S}})^{\frac{1}{2}}[(4+3e^{2}_{{\rm
S}})\sin{g_{{\rm T}}}+6(x_{{\rm S}}y_{{\rm S}}\cos{g_{{\rm
T}}}+y_{{\rm S}}^{2}\sin{g_{{\rm T}}})]-[\frac{(1-e^{2}_{{\rm
S}})^{\frac{1}{2}}}{(1-e^{2}_{{\rm
T}})^{\frac{3}{2}}}+\frac{25}{8}\gamma\frac{3+2e^{2}_{{\rm
T}}}{(1-e^{2}_{{\rm T}})^{3}}(1-\nonumber\\
& & -\frac{3}{2}e^{2}_{{\rm S}})]y_{{\rm
S}}+\frac{15}{8}\gamma\frac{e^{2}_{{\rm T}}}{(1-e^{2}_{{\rm
T}})^{3}}[y_{{\rm S}}\cos{2g_{{\rm T}}}-x_{{\rm S}}\sin{2g_{{\rm
T}}}-\frac{y_{{\rm S}}}{2}(x^{2}_{{\rm
S}}+3y^{2}_{{\rm S}})\cos{2g_{{\rm T}}}+x_{{\rm S}}(x^{2}_{{\rm S}}+2y^{2}_{{\rm S}})\sin{2g_{{\rm T}}}]\\
\frac{{\rm d}y_{{\rm S}}}{{\rm d}\tau} & =
&-\frac{5}{16}\alpha\frac{e_{{\rm T}}}{(1-e^{2}_{{\rm
T}})^\frac{5}{2}}(1-e^{2}_{{\rm S}})^{\frac{1}{2}}[(4+3e^{2}_{{\rm
S}})\cos{g_{{\rm T}}}+6(x_{{\rm S}}y_{{\rm S}}\sin{g_{{\rm
T}}}+x_{{\rm S}}^{2}\cos{g_{{\rm T}}})]+[\frac{(1-e^{2}_{{\rm
S}})^{\frac{1}{2}}}{(1-e^{2}_{{\rm
T}})^{\frac{3}{2}}}+\frac{25}{8}\gamma\frac{3+2e^{2}_{{\rm
T}}}{(1-e^{2}_{{\rm T}})^{3}}(1-\nonumber\\
& & -\frac{3}{2}e^{2}_{{\rm S}})]x_{{\rm
S}}+\frac{15}{8}\gamma\frac{e^{2}_{{\rm T}}}{(1-e^{2}_{{\rm
T}})^{3}}[x_{{\rm S}}\cos{2g_{{\rm T}}}+y_{{\rm S}}\sin{2g_{{\rm
T}}}-\frac{x_{{\rm S}}}{2}(y^{2}_{{\rm
S}}+3x^{2}_{{\rm S}})\cos{2g_{{\rm T}}}-y_{{\rm S}}(y^{2}_{{\rm S}}+2x^{2}_{{\rm S}})\sin{2g_{{\rm T}}}]\\
\frac{{\rm d}g_{{\rm T}}}{{\rm d}\tau} & = & \frac{\beta
(2+3e^{2}_{{\rm S}})}{2(1-e^{2}_{{\rm
T}})^{2}}-\frac{5}{16}\frac{\alpha \beta (1+4e^{2}_{{\rm
T}})}{e_{{\rm T}}(1-e^{2}_{{\rm T}})^{3}}(4+3e^{2}_{{\rm
S}})(x_{{\rm S}}\cos{g_{{\rm T}}}+y_{{\rm S}}\sin{g_{{\rm
T}}})+\frac{5}{8}\beta\gamma\frac{(1-e^{2}_{{\rm
S}})^{\frac{1}{2}}}{(1-e^{2}_{{\rm
T}})^{\frac{7}{2}}}[5e^{2}_{{\rm S}}(11+4e^{2}_{{\rm
T}})+3(1+2e^{2}_{{\rm T}})\times\nonumber\\
& & \times((x^{2}_{{\rm S}}-y^{2}_{{\rm S}})\cos{2g_{{\rm
T}}}+2x_{{\rm S}}y_{{\rm S}}\sin{2g_{{\rm T}}})]\label{gtd}\\
\frac{{\rm d}e_{{\rm T}}}{{\rm d}\tau} & = &
\frac{5}{16}\frac{\alpha \beta}{(1-e^{2}_{{\rm
T}})^{2}}(4+3e^{2}_{{\rm S}})(y_{{\rm S}}\cos{g_{{\rm T}}}-x_{{\rm
S}}\sin{g_{{\rm T}}})-\frac{15}{8}\beta\gamma\frac{e_{{\rm
T}}(1-e^{2}_{{\rm S}})^{\frac{1}{2}}}{(1-e^{2}_{{\rm
T}})^{\frac{5}{2}}}(2x_{{\rm S}}y_{{\rm S}}\cos{2g_{{\rm
T}}}-(x^{2}_{{\rm S}}-y^{2}_{{\rm S}})\sin{2g_{{\rm T}}})
\end{eqnarray}
where
\begin{displaymath}
x_{{\rm S}}=e_{{\rm S}}\cos{g_{{\rm S}}},\hspace{0.5cm}y_{{\rm
S}}=e_{{\rm S}}\sin{g_{{\rm S}}},
\end{displaymath}
\begin{displaymath}
\alpha =\frac{m_{1}-m_{2}}{m_{1}+m_{2}}\frac{a_{{\rm S}}}{a_{{\rm
T}}},\hspace{0.2cm}\beta
=\frac{m_{1}m_{2}M^{\frac{1}{2}}}{m_{3}(m_{1}+m_{2})^{\frac{3}{2}}}(\frac{a_{{\rm
S}}}{a_{{\rm
T}}})^{\frac{1}{2}},\hspace{0.2cm}\gamma=\frac{m_{3}}{M^{\frac{1}{2}}(m_{1}+m_{2})^{\frac{1}{2}}}(\frac{a_{{\rm
S}}}{a_{{\rm T}}})^{\frac{3}{2}}\hspace{0.5cm}
\mbox{and}\hspace{0.5 cm} {\rm
d}\tau=\frac{3}{4}\frac{G^{\frac{1}{2}}m_{3}a^{\frac{3}{2}}_{{\rm
S}}}{a^{3}_{{\rm T}}(m_{1}+m_{2})^{\frac{1}{2}}}{\rm d}t.
\end{displaymath}

In Georgakarakos (2002), we assumed that ${e_{{\rm T}}}$ remained
constant, we neglected terms of order ${e^{2}_{{\rm S}}}$ and
${e^{2}_{{\rm T}}}$ and only the term proportional to ${\beta}$
was retained in equation (\ref{gtd}).  Under those assumptions,
the system was reduced to one that had the following form:

\begin{eqnarray}
\frac{{\rm d}x_{{\rm S}}}{{\rm d}\tau} & = & -By_{{\rm S}}+C\sin{g_{{\rm T}}}\nonumber\\
\frac{{\rm d}y_{{\rm S}}}{{\rm d}\tau} & = & Bx_{{\rm
S}}-C\cos{g_{{\rm T}}}
\label{dior}\\
\frac{{\rm d}g_{{\rm T}}}{{\rm d}\tau} & = & A\nonumber,
\end{eqnarray}
where ${A,B,}$ and ${C}$ were constants.

The solution of the above system was singular for ${A-B=0}$ (this
is to be expected, as the above system of differential equations
corresponds to the motion of  a forced harmonic oscillator with a
constant forcing frequency ${A}$ and a natural frequency ${B}$).
As a consequence, we got an overestimate of the inner eccentricity
for the parameter values that satisfied ${A-B \approx 0}$.  The
source of the problem was the assumption that the outer pericentre
frequency was almost constant, or in other words, the fact that we
only kept the dominant term in the approximate expression for
${\dot{g}_{{\rm T}}}$.  In fact, although the term proportional to
${\alpha\beta}$ in the equation for the outer pericentre is
insignificant compared to the leading term in most of the cases,
there are systems (the systems for which ${A-B \approx 0}$) for
which the ${\alpha\beta}$ term becomes comparable to the leading
term, and hence, its inclusion in the relevant equation is
necessary ( note that ${\dot{e}_{{\rm T}}}$ is also proportional
to ${\alpha\beta}$ to leading order).

In order to deal with the extra term in the equation for the outer
pericentre, we introduce the following two variables (as we did
for the inner binary):

\begin{displaymath}
x_{{\rm T}}=e_{{\rm T}}\cos{g_{{\rm T}}},\hspace{0.5cm}y_{{\rm
T}}=e_{{\rm T}}\sin{g_{{\rm T}}}.
\end{displaymath}
By using equations (12) and (13) and without making any assumption
about the outer binary pericentre and the two eccentricities, we
obtain expressions for ${\dot{x}_{{\rm T}}}$ and ${\dot{y}_{{\rm
T}}}$.  Then, if we neglect terms of order ${e^{2}_{{\rm S}}}$ and
${e^{2}_{{\rm T}}}$ (keep in mind that we deal with initially
circular orbits), we obtain the following system of differential
equations:

\begin{eqnarray}
\frac{{\rm d}x_{{\rm S}}}{{\rm d}\tau} & = & -By_{{\rm S}}+Cy_{{\rm T}}\nonumber\\
\frac{{\rm d}y_{{\rm S}}}{{\rm d}\tau} & = & Bx_{{\rm S}}-Cx_{{\rm
T}}\nonumber\\
\frac{{\rm d}x_{{\rm T}}}{{\rm d}\tau} & = & CDy_{{\rm S}}-Dy_{{\rm T}}\label{dior1}\\
\frac{{\rm d}y_{{\rm T}}}{{\rm d}\tau} & = & -CDx_{{\rm
S}}+Dx_{{\rm T}}\nonumber,
\end{eqnarray}
where
\begin{displaymath}
B=1+\frac{75}{8}\gamma, \hspace{0.2cm} C=\frac{5}{4}\alpha
\hspace{0.2cm} \mbox{and} \hspace{0.1cm} D=\beta.
\end{displaymath}
Note that ${C}$ is slightly different from the one in system
(\ref{dior})(${C=(5/4)\alpha e_{{\rm T}}}$ in system
(\ref{dior})).

System (\ref{dior1}) is a system of four first order linear ordinary
differential equations. The characteristic equation of the system
is
\begin{equation}
\lambda^{4}+(D^{2}+B^{2}+2C^{2}D)\lambda^{2}+(DB-C^{2}D)^{2}=0
\end{equation}
and it has the following four imaginary eigenvalues:
\begin{displaymath}
\lambda_{1,2}=\pm ik_{3}, \hspace{0.1cm} \lambda_{3,4}=\pm ik_{4},
\end{displaymath}
where
\begin{displaymath}
k_{3}=\sqrt{\frac{k_{1}+k_{2}}{2}},\hspace{0.1cm}
k_{4}=\sqrt{\frac{k_{1}-k_{2}}{2}},\hspace{0.1cm}
k_{1}=D^{2}+B^{2}+2C^{2}D \hspace{0.1cm}\mbox{and}\hspace{0.1cm}
k_{2}=\sqrt{(D+B)^{2}[(D-B)^{2}+4C^{2}D]}.
\end{displaymath}
The corresponding eigenvectors are
\begin{displaymath}
\xi_{i}=[\xi_{1i},\xi_{2i},\xi_{3i},1]^{T}, \hspace{1cm} i=1,2,3,4
\end{displaymath}
with
\begin{displaymath}
\xi_{1i}=\frac{C}{\lambda_{i}}\frac{C^{2}D+\lambda^{2}_{i}-BD}{B^2+C^2D+\lambda^{2}_{i}},\hspace{0.1cm}
\xi_{2i}=\frac{C(B+D)}{B^2+C^2D+\lambda^{2}_{i}},\hspace{0.1cm}
\xi_{3i}=\frac{1}{\lambda_{i}}\frac{BC^{2}D-(B^{2}+\lambda^{2}_{i})D}{B^2+C^2D+\lambda^{2}_{i}}.
\end{displaymath}

Now, since we have found the eigenvalues and the corresponding
eigenvectors, we are able to obtain the general solution of system
(\ref{dior1}), which is the following:
\begin{equation}
\bmath{X}=C_{1}\bmath{U_{13}}+C_{2}\bmath{U_{23}}+C_{3}\bmath{U_{14}}+C_{4}\bmath{U_{24}},
\label{secf}
\end{equation}
where
\begin{displaymath} \bmath{X}=[x_{{\rm S}},y_{{\rm
S}},x_{{\rm T}},y_{{\rm T}}]^{T},
\end{displaymath}
\begin{displaymath}
\bmath{U_{1j}}=[F_{1j}\sin{k_{j}\tau},F_{2j}\cos{k_{j}\tau},F_{3j}\sin{k_{j}\tau},\cos{k_{j}\tau}]^{T},
\end{displaymath}
\begin{displaymath}
\bmath{U_{2j}}=[-F_{1j}\cos{k_{j}\tau},F_{2j}\sin{k_{j}\tau},-F_{3j}\cos{k_{j}\tau},\sin{k_{j}\tau}]^{T},
\hspace{0.2cm}\mbox{with}\hspace{0.1cm} j=3,4.
\end{displaymath}
The ${F}$ coefficients are
\begin{displaymath}
F_{1j}=\frac{C}{k_{j}}\frac{C^{2}D-k^{2}_{j}-BD}{B^2+C^2D-k^{2}_{j}},\hspace{0.1cm}
F_{2j}=\frac{C(B+D)}{B^2+C^2D-k^{2}_{j}},\hspace{0.1cm}
F_{3j}=\frac{1}{k_{j}}\frac{BC^{2}D-(B^{2}-k^{2}_{j})D}{B^2+C^2D-k^{2}_{j}},
\end{displaymath}
and finally, the ${C_{i}}$ constants, for initially circular
orbits, are found to be (see second from the end paragraph of section \ref{sss}):
\begin{displaymath}
C_{1}=\frac{1}{F_{24}-F_{23}}(y_{10}-y_{20}F_{24}),\hspace{0.5cm}
C_{2}=\frac{1}{F_{13}F_{34
}-F_{33}F_{14}}(x_{10}F_{34}-x_{20}F_{14}),
\end{displaymath}
\begin{displaymath}
C_{3}=\frac{1}{F_{24}-F_{23}}(y_{20}F_{23}-y_{10}),\hspace{0.5cm}\mbox{and}\hspace{0.5cm}
C_{4}=\frac{1}{F_{13}F_{34
}-F_{33}F_{14}}(x_{20}F_{13}-x_{10}F_{33}),
\end{displaymath}
where ${x_{10}=x_{1}(0)}$, ${y_{10}=y_{1}(0)}$,
${x_{20}=x_{2}(0)}$ and ${y_{20}=y_{2}(0)}$.

\subsection{The new formula} \label{s4}

Now, if we combine the short period solution as given in
subsection (2.1) and the new secular solution we derived in the
previous section (this is done the usual way, i.e. by replacing
the constants in equations (\ref{e11}) and (\ref{e12}) with the
equations for ${x_{{\rm S}}}$ and ${x_{{\rm T}}}$ obtained from
(\ref{secf}), as the latter evolve on a much larger time-scale
than the short period terms and they can practically be treated as
constants for time-scales comparable to the inner and outer
orbital period) and if we average over time and over the initial
phase ${\phi}$, we get:

\begin{eqnarray}
\overline{e_{{\rm in}}^{2}}& = &\frac{m_{3}^{2}}
{M^{2}}\frac{1}{X^{4}}\left(\frac{225}{256}m^{2}_{*}X^{\frac{2}{3}}+\frac{43}{8}+
\frac{61}{3}\frac{1}{X}+\frac{8361}{4096}m^{2}_{*}\frac{1}{X^{\frac{4}{3}}}+\frac{365}{18}\frac{1}{X^{2}}
\right)+\nonumber\\
& & +\frac{1}{2}\left[(F^{2}_{13}+F^{2}_{23})(C^{2}_{1\rm
av}+C^{2}_{2\rm av})+(F^{2}_{14}+F^{2}_{24})(C^{2}_{3\rm
av}+C^{2}_{4\rm av})\right], \label{final}
\end{eqnarray}
where
\begin{eqnarray}
C^{2}_{1\rm av} & = & \frac{1}{(F_{24}-F_{23})^{2}}\left
[(\frac{m_{3}}{M})^{2}\frac{1}{X^{4}}\left(\frac{225}{512}m^{2}_{*}X^{\frac{2}{3}}+2-\frac{45}{512}\frac{m^{2}_{*}}
{X^{\frac{1}{3}}}+\frac{26}{3}\frac{1}{X}+\frac{5661}{8192}\frac{m^{2}_{*}}{X^{\frac{4}{3}}}+\frac{169}{18}
\frac{1}{X^{2}}\right)\right.+\nonumber\\
 & & \left.+F_{24}\frac{m_{3}m_{**}}{M}\frac{1}{X^{\frac{10}{3}}}\left(-\frac{45}{64}m_{*}X^{\frac{1}{3}}-\frac{27}{256}
 \frac{m_{*}}{X^{\frac{2}{3}}}-\frac{1557}{1024}\frac{m_{*}}{X^{\frac{5}{3}}}\right)
 +F^{2}_{24}\frac{m^{2}_{**}}{X^{\frac{8}{3}}}\left(\frac{9}{32}+\frac{9}{64}\frac{1}{X}+\frac{225}{256}
 \frac{1}{X^{2}}\right)\right]\nonumber\\
C^{2}_{2\rm av} & = &
\frac{1}{(F_{13}F_{34}-F_{33}F_{14})^{2}}\left
[(\frac{m_{3}}{M})^{2}\frac{1}{X^{4}}\left(\frac{225}{512}m^{2}_{*}X^{\frac{2}{3}}+\frac{27}{8}+\frac{45}{512}\frac{m^{2}_{*}}
{X^{\frac{1}{3}}}+\frac{35}{3}\frac{1}{X}+\frac{11061}{8192}\frac{m^{2}_{*}}{X^{\frac{4}{3}}}+\frac{98}{9}
\frac{1}{X^{2}}\right)\right.+\nonumber\\
 & & \left.+F_{14}F_{34}\frac{m_{3}m_{**}}{M}\frac{1}{X^{\frac{10}{3}}}\left(-\frac{45}{64}m_{*}X^{\frac{1}{3}}+\frac{27}{256}
 \frac{m_{*}}{X^{\frac{2}{3}}}-\frac{2187}{1024}\frac{m_{*}}{X^{\frac{5}{3}}}\right)
 +F^{2}_{14}\frac{m^{2}_{**}}{X^{\frac{8}{3}}}\left(\frac{9}{32}-\frac{9}{64}\frac{1}{X}+\frac{225}{256}
 \frac{1}{X^{2}}\right)\right]\nonumber\\
 C^{2}_{3\rm av} & = & \frac{1}{(F_{24}-F_{23})^{2}}\left
[(\frac{m_{3}}{M})^{2}\frac{1}{X^{4}}\left(\frac{225}{512}m^{2}_{*}X^{\frac{2}{3}}+2-\frac{45}{512}\frac{m^{2}_{*}}
{X^{\frac{1}{3}}}+\frac{26}{3}\frac{1}{X}+\frac{5661}{8192}\frac{m^{2}_{*}}{X^{\frac{4}{3}}}+\frac{169}{18}
\frac{1}{X^{2}}\right)\right.+\nonumber\\
 & & \left.+F_{23}\frac{m_{3}m_{**}}{M}\frac{1}{X^{\frac{10}{3}}}\left(-\frac{45}{64}m_{*}X^{\frac{1}{3}}-\frac{27}{256}
 \frac{m_{*}}{X^{\frac{2}{3}}}-\frac{1557}{1024}\frac{m_{*}}{X^{\frac{5}{3}}}\right)
 +F^{2}_{23}\frac{m^{2}_{**}}{X^{\frac{8}{3}}}\left(\frac{9}{32}+\frac{9}{64}\frac{1}{X}+\frac{225}{256}
 \frac{1}{X^{2}}\right)\right]\nonumber\\
C^{2}_{4 \rm av} & = &
\frac{1}{(F_{13}F_{34}-F_{33}F_{14})^{2}}\left
[(\frac{m_{3}}{M})^{2}\frac{1}{X^{4}}\left(\frac{225}{512}m^{2}_{*}X^{\frac{2}{3}}+\frac{27}{8}+\frac{45}{512}\frac{m^{2}_{*}}
{X^{\frac{1}{3}}}+\frac{35}{3}\frac{1}{X}+\frac{11061}{8192}\frac{m^{2}_{*}}{X^{\frac{4}{3}}}+\frac{98}{9}
\frac{1}{X^{2}}\right)\right.+\nonumber\\
 & & \left.+F_{13}F_{33}\frac{m_{3}m_{**}}{M}\frac{1}{X^{\frac{10}{3}}}\left(-\frac{45}{64}m_{*}X^{\frac{1}{3}}+\frac{27}{256}
 \frac{m_{*}}{X^{\frac{2}{3}}}-\frac{2187}{1024}\frac{m_{*}}{X^{\frac{5}{3}}}\right)
 +F^{2}_{13}\frac{m^{2}_{**}}{X^{\frac{8}{3}}}\left(\frac{9}{32}-\frac{9}{64}\frac{1}{X}+\frac{225}{256}
 \frac{1}{X^{2}}\right)\right]\nonumber
\end{eqnarray}
and
\begin{displaymath}
m_{**}=\frac{m_{1}m_{2}}{(m_{1}+m_{2})^{\frac{4}{3}}M^{\frac{2}{3}}}.
\end{displaymath}

\section{NUMERICAL TESTING}
\label{s5}

In order to test the validity of the formulae derived in the
previous papers, we integrated the full equations of motion
numerically, using a symplectic integrator with time
transformation (Mikkola 1997).

The code calculates the relative position and velocity vectors of
the two binaries at every time step.  Then, by using standard two
body formulae, we computed the orbital elements of the two
binaries. The various parameters used by the code, were given the
following values: writing index ${Iwr=1}$, method coefficients
${a1=1}$ and ${a2=15}$, correction index ${icor=1}$. The average
number of steps per inner binary period ${NS}$, was given the
value of ${60}$.

For our simulations, we use  the two mass ratios we defined in
Georgakarakos (2006), i.e.
\begin{displaymath}
K1=\frac{m_{1}}{m_{1}+m_{2}} \hspace{0.2cm}\mbox{and}
\hspace{0.2cm} K2=\frac{m_{3}}{m_{1}+m_{2}},
\end{displaymath}
with ${0.001 \leq K1 \leq 0.5}$ and ${0.001 \leq K2 \leq 1000}$
(although our main interest is for systems with comparable masses,
we also test the theory for some planetary mass ratios too), but
we studied a few more pairs than in Georgakarakos (2006). We also
used units such that ${G=1}$ and ${m_{1}+m_{2}=1}$ and we always
started the integrations with ${a_{1}=1}$.  In that system of
units, the initial conditions for the numerical integrations were
as follows:

\begin{displaymath}
r_{1}=1,\hspace{0.5cm} r_{2}=0,\hspace{0.5cm}
R_{1}=a_{2}\cos{\phi},\hspace{0.5cm}
R_{2}=a_{2}\sin{\phi},\hspace{0.5cm}
\end{displaymath}
\begin{displaymath}
\dot{r}_{1}=0,\hspace{0.5cm} \dot{r}_{2}=1,\hspace{0.5cm}
\dot{R}_{1}=-\sqrt{\frac{M}{a_{2}}}\sin{\phi},\hspace{0.5cm}
\dot{R}_{2}=\sqrt{\frac{M}{a_{2}}}\cos{\phi}. \hspace{0.5cm}
\end{displaymath}

\subsection{Short period evolution}

First, we present some results from testing equations (\ref{e11})
and (\ref{e12}). As in all our previous papers, we compare the
averaged over time numerical and theoretical inner eccentricity.
In addition to that, we also compare the averaged over time and
initial phase numerical and theoretical inner eccentricity in
order to get a picture of the error dependence on ${\phi}$.

Generally, the error  gets larger for increasing values of
${K_{2}}$ and, for fixed mass ratios, it goes down as the period
ratio increases.  Also, the error for a specific ${K_{2}}$ seems
to be almost independent from the variation of ${K_{1}}$ (for a
specific value of ${\phi}$).  An interesting finding was, that for
some systems, the error was seriously affected by the choice of
the initial phase ${\phi}$, especially for systems with larger
${K_{2}}$, i.e. large ${m_{3}}$. For instance,  for
${K_{2}=1000}$, ${X=10}$  and ${\phi=90^{\circ}}$, the error is
around ${28\%}$; however, that error can even vanish for certain
values of the initial phase. Since the formulae depend on
${\phi}$, we expected the error to depend on the initial phase
too, but that result was a bit surprising.  Figure 1 is a plot of
the percentage error ${p}$ against the initial phase ${\phi}$ for
${K_{1}=0.001}$ and ${K_{2}=1000}$. The shape and size of the
curves remain almost the same as ${K_{1}}$ varies. The graphs are
also similar when we plot them for smaller ${K_{2}}$. However, as
${K_{2}}$ gets smaller, the amplitude of the oscillation of the
error becomes smaller too, and for very small outer masses
(${K_{2} <0.1}$), the curves are almost flat.

Figures 2 show the percentage error between the averaged (top row
graphs: averaged over time inner eccentricity with
${\phi=90^{\circ}}$; bottom row graphs: averaged over time and
initial phase inner eccentricity) numerical and theoretical inner
eccentricity.  The left graphs are for ${0.001\leq K_{2} \leq 1}$
while the right graphs are for ${0.001\leq K_{2} \leq 50}$.  For
${K_{2} > 50}$, the error remains almost the same as for
${K_{2}=50}$.  All graphs are for ${K_{1}=0.05}$.  For other
values of ${K_{1}}$, we get similar graphs, as the error does not
vary much for the rest of the values of ${K_{1}}$.

Finally, simulations that were performed including the ${P_{51}}$
terms in  equations (\ref{e11}) and (\ref{e12}), showed an extra
improvement of around ${1\%}$.

\begin{figure}
\begin{center}
\includegraphics[width=80mm,height=60mm]{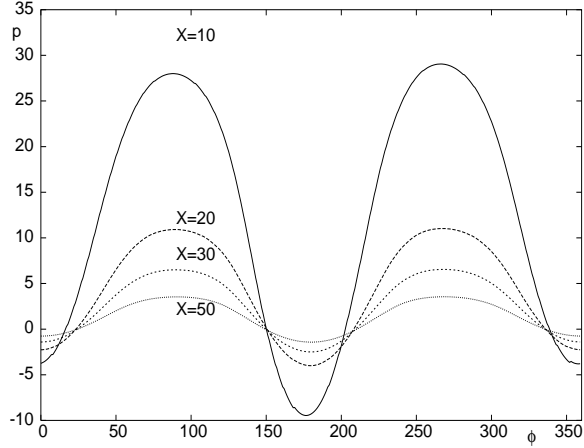}
\caption[]{Percentage error ${p}$ against initial phase ${\phi}$
for ${K_{1}=0.001}$ and ${K_{2}=1000}$.}
\end{center}
\end{figure}

\begin{figure}
\begin{center}
\includegraphics[width=80mm,height=60mm]{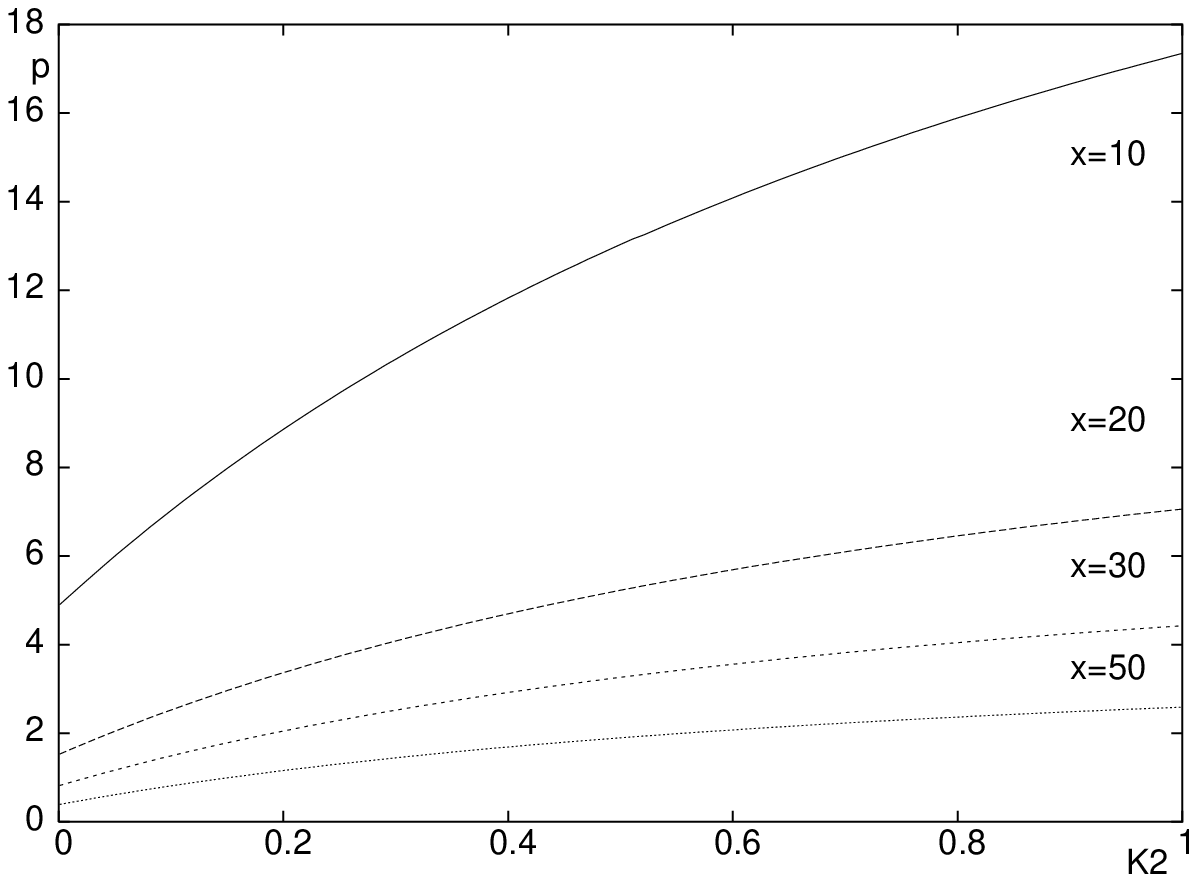}
\includegraphics[width=80mm,height=60mm]{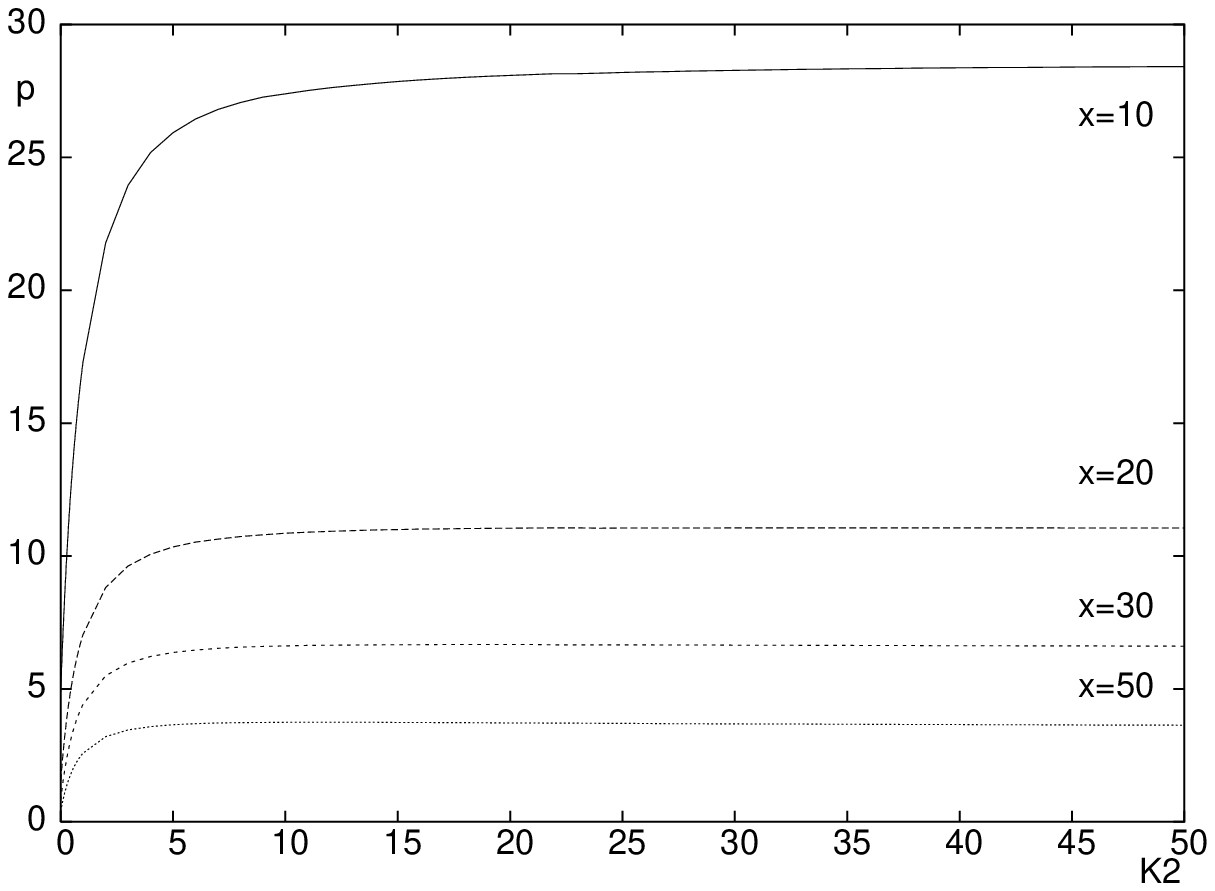}
\includegraphics[width=80mm,height=60mm]{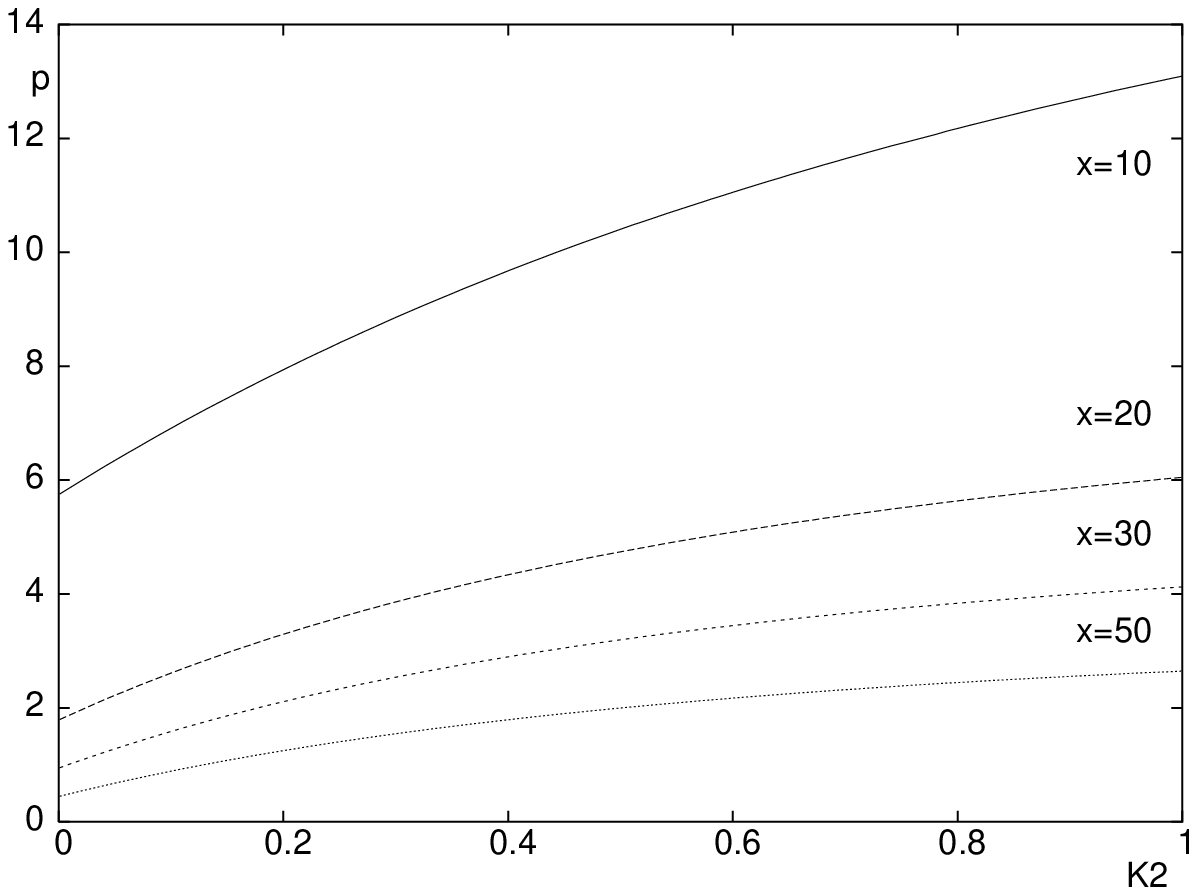}
\includegraphics[width=80mm,height=60mm]{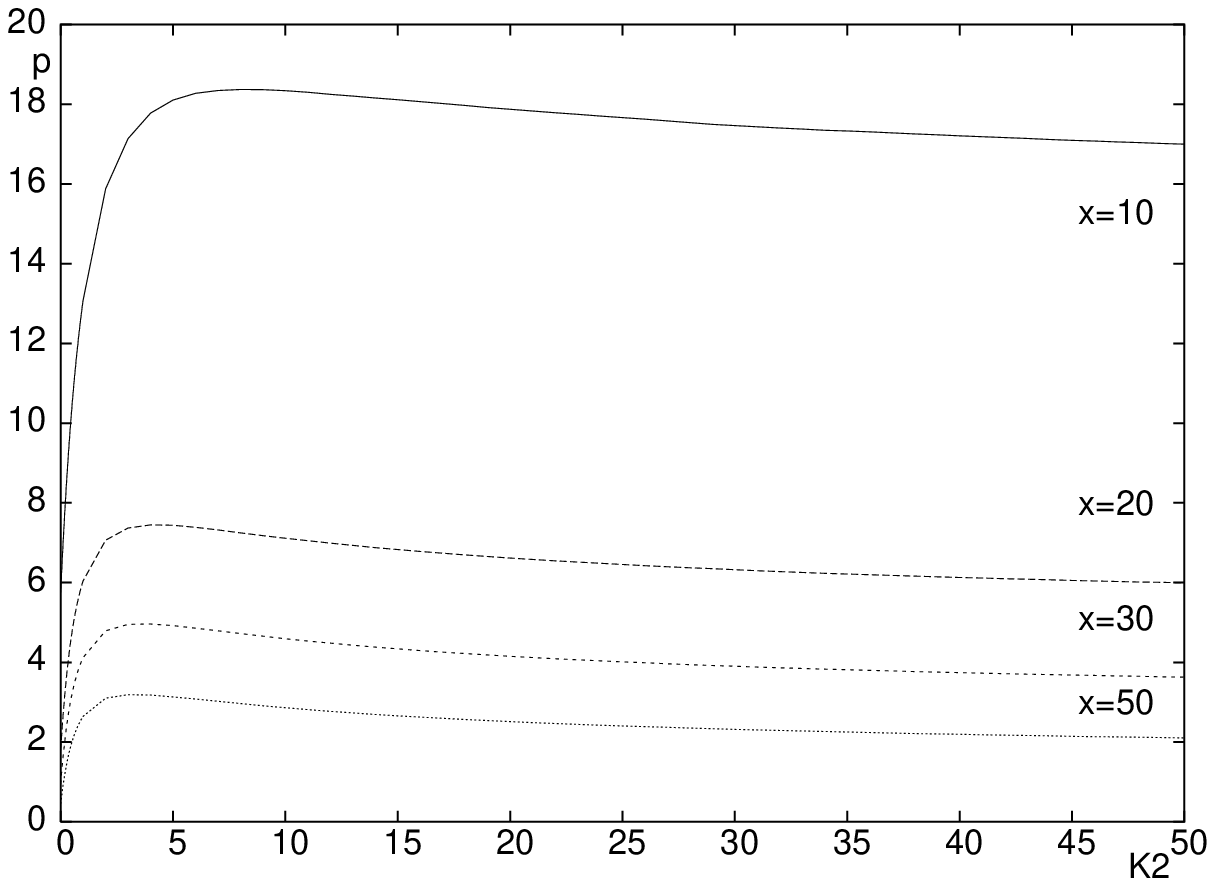}
\caption[]{Percentage error between the averaged short period
numerical and theoretical eccentricity against ${K_{2}}$. The top
row graphs refer to the averaged over time inner eccentricity
(${\phi=90^{\circ}}$) , while the bottom row graphs refer to the
averaged over time and initial phase inner eccentricity.  In all
cases, ${K_{1}=0.05}$.}
\end{center}
\end{figure}

\subsection{Long period evolution}

In this section, we present the results from comparing equation
(\ref{final}) with the numerical solution of the full equations of
motion.

Each triple system was integrated numerically for
${\phi=0^{\circ}-360^{\circ}}$ with a step of ${45^{\circ}}$.
After each run, ${e^{2}_{{\rm in}}}$ was averaged over time using
the trapezium rule and after the integrations for all ${\phi}$
finished, we averaged over ${\phi}$ by using the rectangle rule.
The integrations were also done for smaller steps in ${\phi}$
(e.g. ${1^{\circ}}$), but there was not any difference in the
outcome.

These results are presented in Table 1, which gives the percentage
error between the averaged numerical ${e^{2}_{{\rm in}}}$ and
equation (\ref{final}).  The error is accompanied by the period of
the oscillation of the eccentricity, which is the same as the
integration time (wherever there is no period given, we integrated
for an outer orbital period, i.e. ${2\pi X}$, as there was not any
noticeable secular evolution). There are four values per
(${K1,K2}$) pair, corresponding, from top to bottom, to
${X=10,20,30,50}$ respectively.

Generally, the new formula seems to work better and to provide
more consistent results than the previous one.  For most systems
that show secular evolution, our model (the averaged square
eccentricity) has an error of about ${10\%}$ for ${X=10}$,
independently of the mass ratios of the system. There are a few
systems with larger error (e.g. for ${K_{2}=0.5}$, but these
systems are dominated by short period evolution) and one with a
small negative value ( ${K_{1}=0.3}$, ${K_{2}=0.1}$ with ${-3\%}$
error). This specific system looks a bit odd in Table 1, but if
there were more systems in the table, the change in the error
would look more smooth (e.g. the error for a system with
${K_{1}=0.2}$ and ${ K_{2}=0.1}$ is ${4.5\%}$). Basically, the
error we get from the short period solution (and which is the main
source of the error in the final formula) is passed to the secular
solution by means of the initial conditions. As a result, our
model has a longer period and some difference in the amplitude is
also noticeable. As ${X}$ increases, the period and amplitude
differences go down and consequently the error goes down too.

Regarding the new secular solution we derived in subsection
(\ref{s3}), it appears that we have achieved our aim, i.e. to
correct the problem that arose from the singularity ${A-B=0}$ in
the old solution.  For example, if one checks the corresponding
table in Georgakarakos (2006), he sees that for ${K_{1}=0.05}$ and
${K_{2}=0.01}$, the errors for ${X=10,20,30,50}$  are
${-1.7,-15.9,-17,-37}$ respectively. Using our new formula, the
corresponding errors become ${9.4,3.9,3,1.6}$. Figures 3
demonstrate the improvement that is achieved by the use of our new
formula.  Both column graphs are plots of inner eccentricity
against time.  The left column graphs are for a system with
${K_{1}=0.3}$, ${K_{2}=0.09}$, ${X=10}$ and ${\phi=90^{\circ}}$,
while the parameter values for the right column graphs are
${K_{1}=0.4}$, ${K_{2}=0.065}$, ${X=50}$ and ${\phi=90^{\circ}}$.
The top graphs come from the numerical integration of the full
equations of motion, the middle graphs are based on the old
theoretical solution of Georgakarakos (2002) and the bottom graphs
are produced by our new equations. The improvement is clearly
noted (period and amplitude of oscillation).  Note that, although
${X=50}$ for the right column graphs, and one may expect not to
have a significant error, the old equations fail badly.

Finally, we would like to mention here, that , after a more
careful inspection of the systems in Table 1 of Georgakarakos
(2006), a few secular periods were corrected and some other
systems that were not given a secular period, they now have one,
as it can be seen in Table 1 of this paper.

\begin{table}
\caption[]{Percentage error between the averaged numerical
${e^{2}_{1}}$ and equation (\ref{final}).  For each (${K1,K2}$) pair, there
are four entries, corresponding, from top to bottom, to
${X=10,20,30,50}$ respectively.  The numbers in parentheses are
the secular periods of the systems.  In
the system of units used, the inner binary period is ${T_{{\rm in}}=2\pi}$.} \vspace{0.1 cm}
\begin{center}
{\small \begin{tabular}{c c c c c c c c c c}\hline
${K2\backslash\ K1}$ & ${0.001}$ & ${0.005}$ & ${0.01}$ & ${0.05}$ & ${0.1}$ & ${0.3}$ & ${0.5}$\\
\hline
0.001 & 11.9 (${1.15\times10^{6}}$) & 9.9 (${4.7\times 10^{5}}$)  & 9.2 (${2\times 10^{5}}$)   & 9.5 (${3.5\times 10^{4}}$)  & 9.8 (${1.9\times 10^{4}}$)  & 11 (${7.5\times 10^{3}}$)   & 12.2  \\
      & 3 (${5\times 10^{6}}$)       & 1.8 (${3.5\times 10^{6}}$)  & 4.3 (${1.1\times 10^{6}}$) & 3.1 (${1.9\times 10^{5}}$)  & 3.1 (${ 10^{5}}$)     & 3.3 (${4.2\times 10^{4}}$)  & 3.7  \\
      & 2.4 (${1.05\times 10^{7}}$)  & 3.7(${1.05\times 10^{7}}$)  & 4.1 (${3\times 10^{6}}$)   & 1.8 (${5\times 10^{5}}$)    & 1.7 (${2.6\times 10^{5}}$)  & 1.7 (${1.1\times 10^{5}}$)  & 1.9  \\
      & -0.5 (${3\times 10^{7}}$)    & 2 (${5\times 10^{7}}$)      & 1.1 (${1.2\times 10^{7}}$) & 0.9 (${1.7\times 10^{6}}$)  & 0.9 (${8.5\times 10^{5}}$)  & 0.8 (${3.7\times 10^{5}}$)  & 0.8  \\
0.005 & 11.4 (${1.65\times 10^{5}}$) & 12.8 (${2.2\times 10^{5}}$) & 10.7 (${3\times 10^{5}}$)  & 9.6 (${4.2\times 10^{4}}$)  & 9.8 (${2.1\times 10^{4}}$)  & 11 (${8.5\times 10^{3}}$)   & 12.2 \\
      & 4 (${6.45\times 10^{5}}$)    & 5.9 (${9\times 10^{5}}$)    & 5.8 (${1.6\times 10^{6}}$) & 3.3 (${2.5\times 10^{5}}$)  & 3.1 (${1.15\times 10^{5}}$) & 3.4 (${4.4\times 10^{4}}$)  & 3.7 \\
      & 2.3 (${1.5\times 10^{6}}$)   & 3.8 (${2\times 10^{6}}$)    & 3.4 (${3.5\times 10^{6}}$) & 1.3 (${7.2\times 10^{5}}$)  & 2 (${3\times 10^{5}}$)      & 1.8 (${1.15\times 10^{5}}$) & 1.9 \\
      & 1.1(${4.3\times 10^{6}}$)    & 1.8 (${5.5\times 10^{6}}$)  & 1.4 (${8.6\times 10^{6}}$  & -0.5 (${2.7\times 10^{6}}$) & 1.4 (${1.03\times 10^{6}}$) & 0.8 (${4\times 10^{5}}$)    & 0.8 \\
0.01  & 11.4 (${8\times 10^{4}}$)    & 11.1 (${10^{5}}$)     & 11 (${1.2\times 10^{5}}$)  & 9.4 (${5.6\times 10^{4}}$)  & 10 (${2.3\times 10^{4}}$)   & 11 (${9\times 10^{3}}$)     & 12.2 \\
      & 4 (${3\times 10^{5}}$)       & 4.7 (${3.6\times 10^{5}}$)  & 3.2 (${5\times 10^{5}}$)   & 3.9 (${3.8\times 10^{5}}$)  & 4.3 (${1.3\times 10^{5}}$)  & 3.4 (${5\times 10^{4}}$)    & 3.8 \\
     & 2.1 (${7.8\times 10^{5}}$)    & 3.1 (${8\times 10^{5}}$)    & 1.4 (${1.1\times 10^{6}}$) & 3 (${1.22\times 10^{6}}$)   & 2.9 (${3.7\times 10^{5}}$)  & 1.5 (${1.45\times 10^{5}}$) & 1.9 \\
     & 1.2 (${2\times 10^{6}}$)      & 0.7 (${2.5\times 10^{6}}$)  & 1.4 (${2.8\times 10^{6}}$) & 1.6(${6\times 10^{6}}$)    & -0.9 (${1.5\times 10^{6}}$) & 0.9 (${4.4\times 10^{5}}$)  & 0.8 \\
0.05 & 12.5                    & 12.5 (${1.4\times 10^{4}}$) & 12.6 (${1.5\times 10^{4}}$)& 12.3 (${2.2\times  10^{4}}$)& 9.2 (${3.5\times 10^{4}}$)  & 12.2 (${1.8\times 10^{4}}$) & 12.5 \\
     & 4.4                     & 4.6 (${6.5\times 10^{4}}$)  & 4.5 (${7\times 10^{4}}$)   & 6.3   (${9\times 10^{4}}$)     & 4.5 (${1.6\times 10^{5}}$)  & 5.4 (${1.2\times 10^{5}}$)  & 3.9 \\
     & 2.4                     & 2.8 (${1.45\times 10^{5}}$) & 2.9 (${1.5\times 10^{5}}$) & 4.4 (${2\times 10^{5}}$)    & 2.5 (${3.4\times 10^{5}}$)  & 2.8 (${4.2\times 10^{5}}$)  & 2 \\
     & 1.1                     & 1.3  (${4.6\times 10^{5}}$) & 1.3 (${4.7 \times10^{5}}$) & 1.7 (${5.7\times 10^{5}}$)  & 2.2 (${8.2\times 10^{5}}$)  & 1.8(${2.6\times 10^{6}}$)   & 0.9 \\
0.1  & 13.9                    & 13.5(${7\times 10^{3}}$)    & 13.2(${9 \times10^{3}}$)   & 13 (${9 \times10^{3}}$)     & 12.2 (${1.1\times 10^{4}}$) & -3 (${3.3\times 10^{4}}$)   & 12.8  \\
     & 5.4                     & 5.3 (${3.4\times 10^{4}}$)  & 5.4 (${3.2\times 10^{4}}$) & 5.2 (${4\times 10^{4}}$)    & 4.5 (${5\times 10^{4}}$)    & 2.2 (${1.25\times 10^{5}}$) & 4 \\
     & 3.3                     & 3.2(${7.9\times 10^{4}}$)   & 3   (${9\times 10^{4}}$)   & 3.4 (${9 \times10^{4}}$)    & 3 (${1.1\times 10^{5}}$)    & 2.4 (${2.35\times 10^{5}}$) & 2.1 \\
     & 1.8                     & 1.8 (${2.2\times 10^{5}}$)  & 1.8 (${2.25\times 10^{5}}$)& 2 (${2.5\times 10^{5}}$)    & 1.5 (${3\times 10^{5}}$)    & 2.1 (${5.2\times 10^{5}}$)  & 1 \\
0.5  & 21.3                    & 21.2                  & 21                   & 19.1 (${1.5\times 10^{3}}$) & 18.3 (${1.5\times 10^{3}}$) & 15.5(${1.8\times 10^{3}}$)  & 15 \\
     & 10.1                    & 9.9                   & 9.7                  & 8.9 (${8.5\times 10^{3}}$)  & 8.4 (${8.5\times 10^{3}}$)  & 6.4 (${8\times 10^{3}}$)    & 5.1 \\
     & 6.8                     & 6.7                   & 6.5                  & 5.9 (${2\times 10^{4}}$)    & 5.6 (${2 \times10^{4}}$)    & 3.9 (${2.2\times 10^{4}}$)  & 2.8 \\
     & 4.2                     & 4.1                   & 3.9                  & 3.8 (${5.7\times 10^{4}}$)  & 3.7 (${5.7\times 10^{4}}$)  & 2.7 (${6.5\times 10^{4}}$)  & 1.3  \\
1    & 26.7                    & 26.6                  & 26.4                 & 25                    & 23.5                  & 18.9                  & 17.3  \\
     & 12.8                    & 12.7                  & 12.6                 & 11.5                  & 10.4(${5 \times10^{3}}$)    & 7.3 (${5\times 10^{3}}$)    & 5.9 \\
     & 8.8                     & 8.7                   & 8.6                  & 7.7 (${1.2\times 10^{4}}$)  & 6.8 (${1.5\times 10^{4}}$)  & 4.5 (${1.3\times 10{4}}$)   & 3.2 \\
     & 5.6                     & 5.5                   & 5.4                  & 4.9 (${2.7\times 10^{4}}$)  & 4.3 (${4.5\times 10^{4}}$)  & 2.6 (${4.2\times 10^{4}}$)  & 1.6 \\
3    & 34.2                    & 34.1                  & 33.9                 & 32.5                  & 28.2 (${1.3\times 10^{3}}$) & 23.2 (${10^{4}}$)     & 22.2 \\
     & 15.6                    & 15.5                  & 15.3                 & 14.5                  & 13.1                  & 9                     & 7.4 \\
     & 10.7                    & 10.6                  & 10.5                 & 9.8                   & 8.7                   & 5.4                   & 4 \\
     & 6.9                     & 6.8                   & 6.8                  & 6.4                   & 5.4                   & 3                     & 2 \\
5    & 35.8                    & 35.7                  & 35.5                 & 34.1                  & 32.3                  & 26.5                  & 24.3 \\
     & 15.6                    & 15.6                  & 15.4                 & 14.5                  & 13.3                  & 9.4                   & 7.9 \\
     & 10.6                    & 10.5                  & 10.4                 & 9.7                   & 8.7                   & 5.6                   & 4.3 \\
     & 6.8                     & 6.8                   & 6.7                  & 6.3                   & 5.5                   & 3.1                   & 2.1 \\
10   & 36                      & 35.9                  & 35.7                 & 34.5                  & 32.9                  & 28.1                  & 26.2 \\
     & 14.8                    & 14.7                  & 14.6                 & 13.8                  & 12.8                  & 9.6                   & 8.4 \\
     & 9.8                     & 9.7                   & 9.6                  & 8.9                   & 8.2                   & 5.6                   & 4.6 \\
     & 6.2                     & 6.2                   & 6.1                  & 5.6                   & 5.4                   & 3.1                   & 2.2 \\
50   & 32.7                    & 32.7                  & 32.6                 & 31.9                  & 31.2                  & 28.9                  & 28.1 \\
     & 11.8                    & 11.8                  & 11.7                 & 11.3                  & 10.8                  & 9.4                   & 8.9 \\
     & 7.2                     & 7.2                   & 7.2                  & 6.8                   & 6.4                   & 5.2                   & 4.8 \\
     & 4.3                     & 4.3                   & 4.2                  & 4                     & 3.7                   & 2.7                   & 2.4 \\
100  & 31.4                    & 31.4                  & 31.4                 & 30.9                  & 30.4                  & 28.9                  & 28.3 \\
     & 10.9                    & 10.8                  & 10.8                 & 10.5                  & 10.2                  & 9.2                   & 8.9 \\
     & 6.5                     & 6.4                   & 6.4                  & 6.2                   & 5.9                   & 5.1                   & 4.8 \\
     & 3.7                     & 3.7                   & 3.6                  & 3.5                   & 3.2                   & 2.6                   & 2.3  \\
500  & 29.7                    & 29.6                  & 29.6                 & 29.5                  & 29.3                  & 28.7                  & 28.5 \\
     & 9.6                     & 9.6                   & 9.6                  & 9.5                   & 9.4                   & 9.1                   & 8.9 \\
     & 5.4                     & 5.4                   & 5.4                  & 5.3                   & 5.2                   & 4.9                   & 4.8 \\
     & 2.9                     & 2.9                   & 2.9                  & 2.8                   & 2.7                   & 2.5                   & 2.4 \\
1000 & 29.3                    & 29.4                  & 29.2                 & 29.1                  & 29                    & 28.7                  & 28.5 \\
     & 9.4                     & 9.4                   & 9.4                  & 9.3                   & 9.2                   & 9                     & 8.9 \\
     & 5.2                     & 5.2                   & 5.2                  & 5.2                   & 5.1                   & 4.9                   & 4.8 \\
     & 2.7                     & 2.7                   & 2.7                  & 2.6                   & 2.6                   & 2.4                   & 2.4 \\

\hline
\end{tabular}}
\end{center}
\end{table}

\begin{figure}
\begin{center}
\includegraphics[width=80mm,height=60mm]{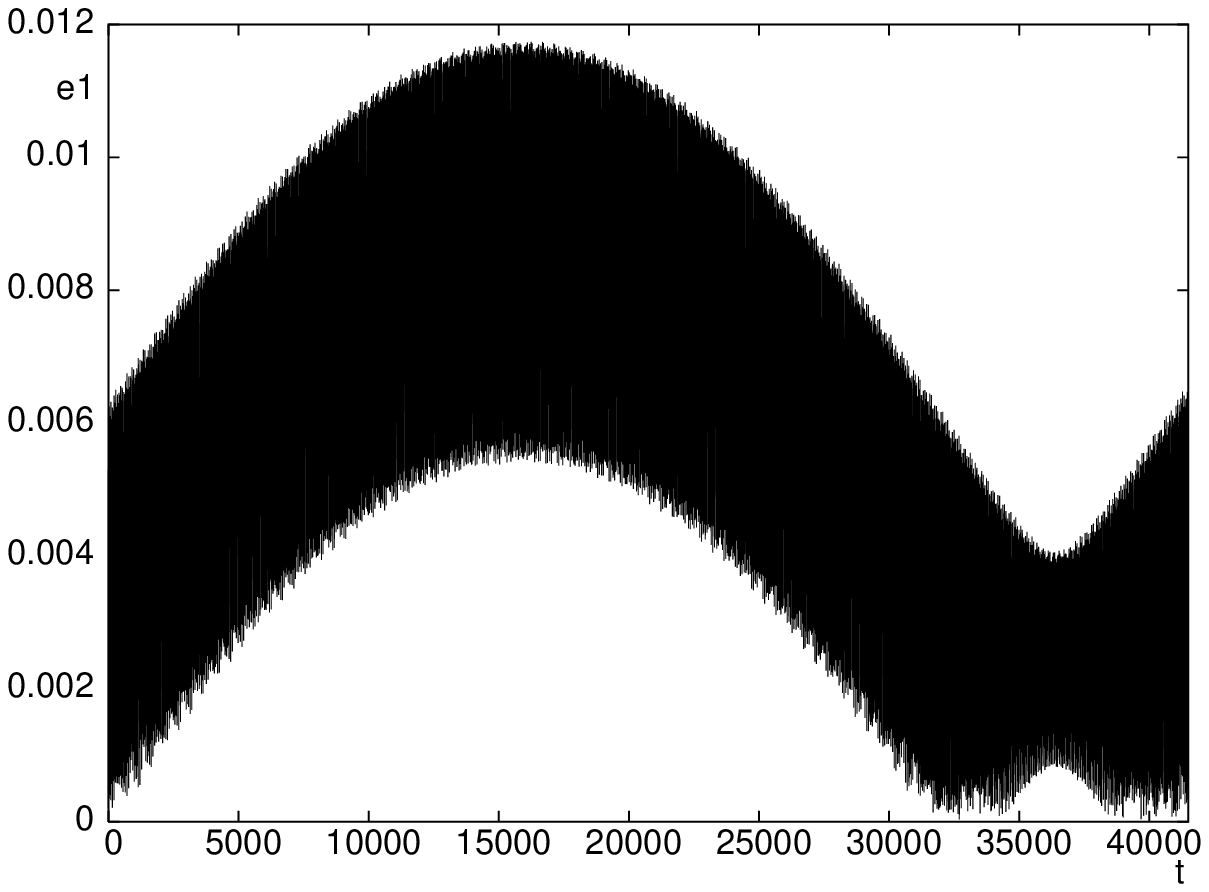}
\includegraphics[width=80mm,height=60mm]{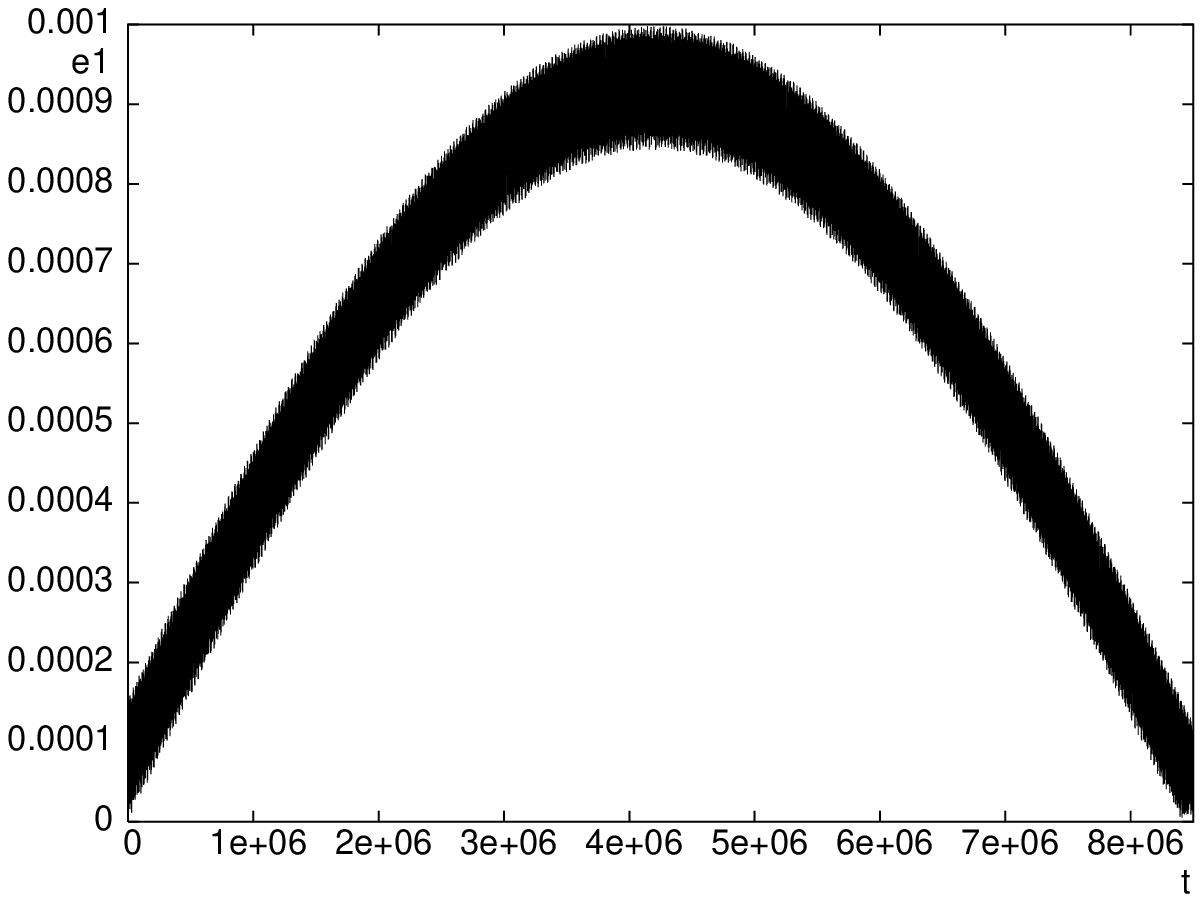}
\includegraphics[width=80mm,height=60mm]{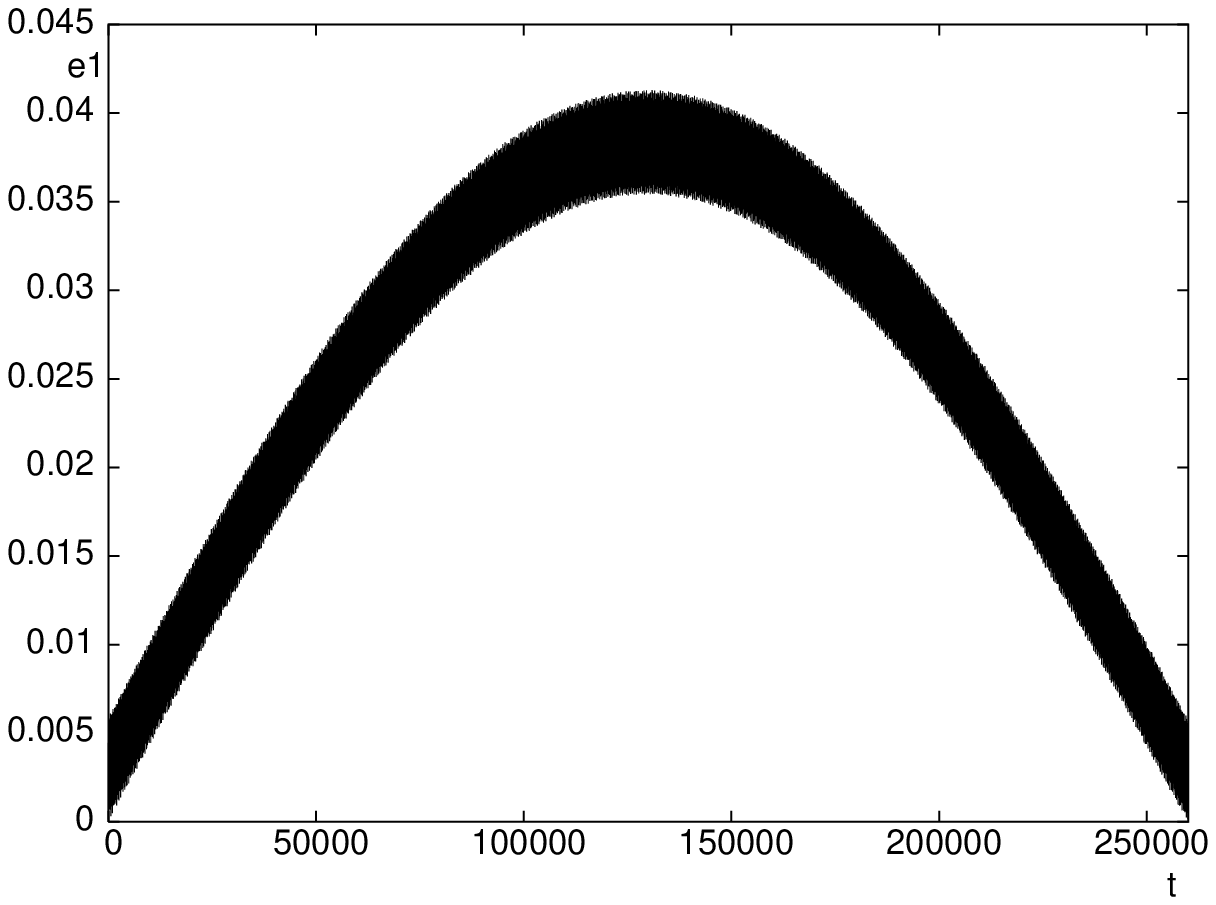}
\includegraphics[width=80mm,height=60mm]{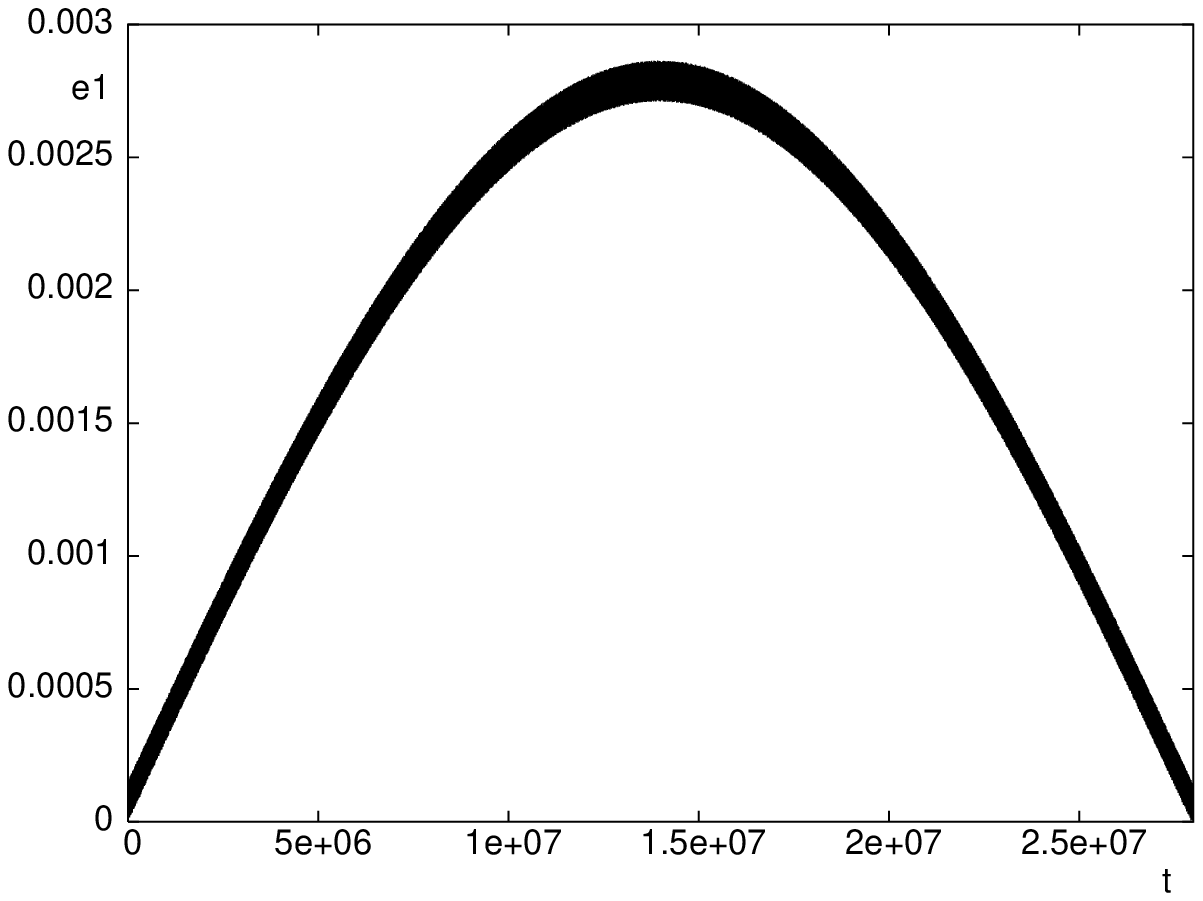}
\includegraphics[width=80mm,height=60mm]{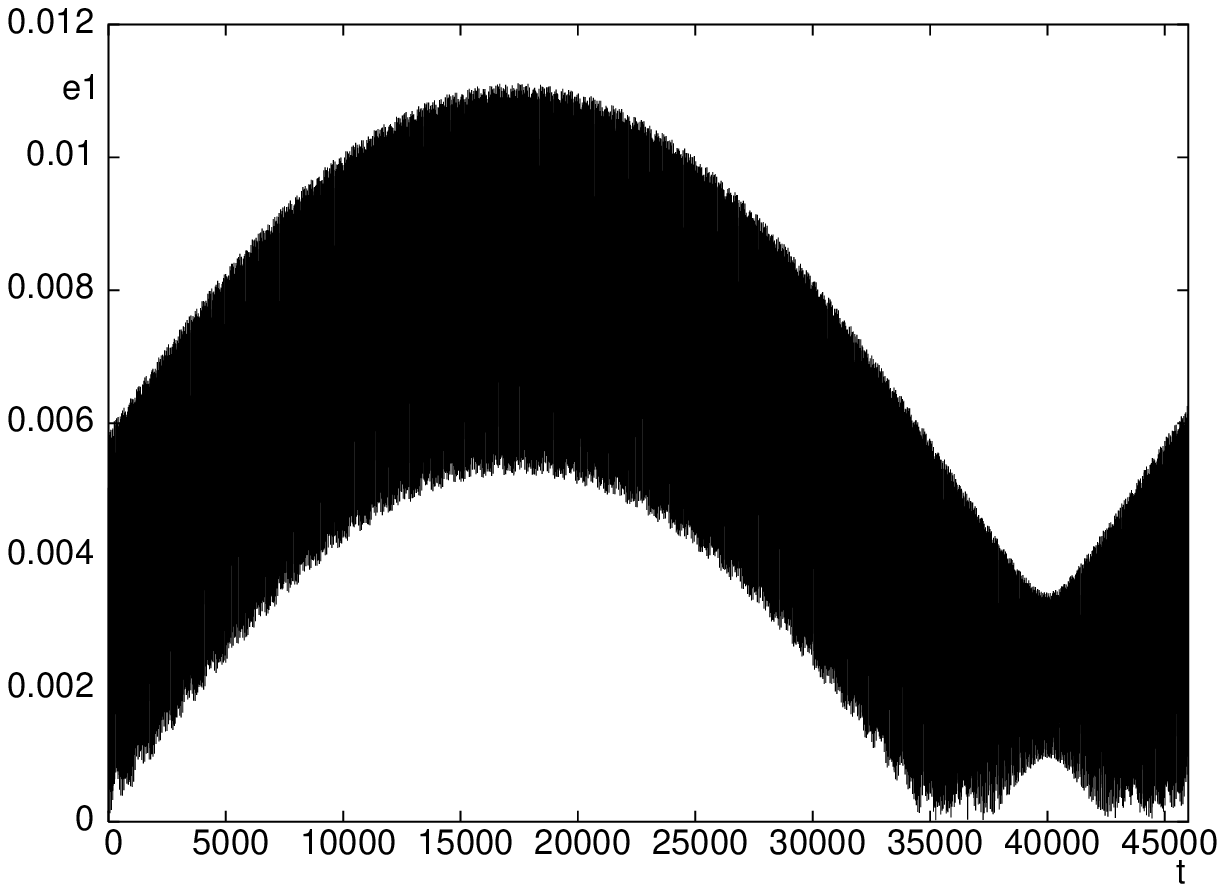}
\includegraphics[width=80mm,height=60mm]{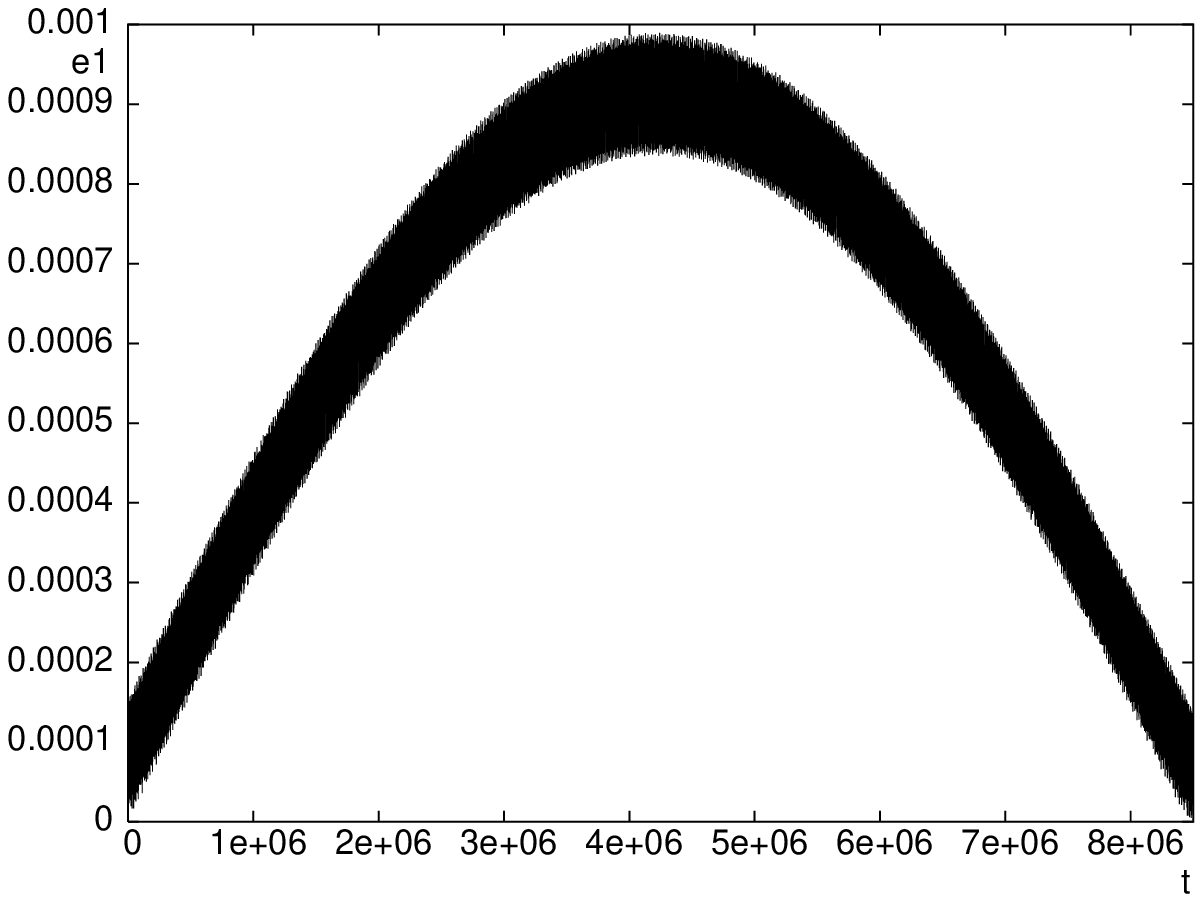}
\caption[]{Inner eccentricity against time.  Left column graphs:
${K_{1}=0.3}$, ${K_{2}=0.09}$, ${X=10}$ and ${\phi=90^{\circ}}$.
Right column graphs: ${K_{1}=0.4}$, ${K_{2}=0.065}$, ${X=50}$ and
${\phi=90^{\circ}}$.The top graphs come from the numerical
integration of the full equations of motion, the middle graphs
show our theoretical solution based on the secular solution of
Georgakarakos (2002) and the bottom graphs are based on the new
secular solution.}
\end{center}
\end{figure}

\section{SUMMARY}

An even distant or light companion is expected to inject
eccentricity to a close binary.  A slight change in the binary
separation can have drastic consequences on its subsequent
evolution.  Therefore, it is  very useful to obtain analytical
expressions about the amount of eccentricity injected into the
binary orbit and the time-scales on which it is done.  In
addition, the analytical expressions could also be used for other
purposes, such as the detection of a companion to a binary system
(Mazeh and Shaham 1979), the determination of orbital elements of
stellar triples or to set constrains to the masses and orbital
elements of exoplanets.

In a series of papers, we derived formulae for estimating the
inner eccentricity in hierarchical triple systems with various
orbital characteristics and with the inner eccentricity being
initially zero. The secular part of the solution we obtained in
those papers, failed for certain combinations of the masses and
the orbital parameters of some systems. In the current paper, we
managed to obtain a secular solution for the inner eccentricity
that had no such weakness.  However the derivation applies only to
hierarchical triple systems with initially circular outer
binaries. For non-circular outer binaries, the system of the
approximate secular differential equations is non-linear and
therefore we can not obtain an improved solution as we did for the
circular case. The numerical integrations of the full equations of
motion confirmed that our new model works well where the previous
one failed to do so.

Our future aim is to see whether a similar solution can be
obtained for the non-coplanar formula of Georgakarakos (2004).

\section*{ACKNOWLEDGMENTS}
The author wants to thank the anonymous referee, whose comments helped to
improve the original manuscript. I also want to thank Douglas Heggie for 
his comments on certain aspects of this paper.

\section*{REFERENCES}

Blaes O., Lee M. H., Socrates A., 2002, ApJ, 578, 775\\
Eggleton P., 2006, Evolutionary Processes in Binary and Multiple
Stars.  Cambridge University Press, UK\\
Eggleton P. P., Kiseleva L. G., 1996, in Wijers R. A. M. J.,
Davies M. B., eds, Proc. NATO Adv. Study Inst., Evolutionary
Processes in Binary Stars.  Kluwer Dordrecht, p. 345\\
Fekel F. C., Jr.; Tomkin J., 1982, ApJ 263, 289\\
Ford E. B., Kozinsky B., Rasio F. A., 2000, ApJ, 535, 385\\
Georgakarakos N., 2002, MNRAS, 337, 559\\
Georgakarakos N., 2003, MNRAS, 345, 340\\
Georgakarakos N., 2004, CeMDA, 89, 63\\
Georgakarakos N., 2006, MNRAS, 366, 566\\
Hinkle K. H., Fekel F. C., Johnson D. S., Scharlach W. W. G., 1993, AJ, 105, 1074\\
Jha S., Torres G., Stefanik R. P., Latham D. W., Mazeh T., 2000, MNRAS, 317, 375\\
Kiseleva-Eggleton L.G., Eggleton P.P., 2001, in Podsiadlowski P.,
Rappaport S., King A. R., D'Antona F., Burderi L., eds, ASP Conf.
Ser. Vol 229, Evolution of Binary and Multiple Star Systems.
Astron. Soc. Pac., San Francisco, p. 91\\
Kiseleva L. G., Eggleton P. P., Mikkola S., 1998, MNRAS, 300, 292\\
Krymolowski Y., Mazeh T., 1999, MNRAS, 304, 720\\
Libert A., Henrard J., 2005, CeMDA, 93, 187\\
Marchal C., 1990, The Three-Body Problem.  Elsevier Science Publishers, the Netherlands\\
Mazeh T., 2008, in Goupil M.-J., Zahn J.-P., eds, EAS Publ. Ser. Vol 29, Tidal Effects in
Stars, Planets and Disks, p. 1\\
Mazeh T.,Shaham J., 1979, A ${\&}$ A, 77, 145\\
Migaszewski C., Gozdziewski K., 2008, MNRAS, 388, 789\\
Mikkola S., 1997, CeMDA, 67, 145\\
Lee M. H., Peale S. J., 2003, ApJ 592, 1201\\
Tokovinin A.A., 1997, A ${\&}$ AS, 124, 75\\

\end{document}